\documentclass[compsoc, conference, a4paper, 10pt, times]{IEEEtran}

\usepackage{mdwmath}
\usepackage{mdwtab}
\usepackage{eqparbox}

\usepackage{fixltx2e}
\usepackage{stfloats}

\usepackage{bibspacing}
\usepackage[pdftex]{graphicx}
\graphicspath{{figures/}}
\DeclareGraphicsExtensions{.jpg,.png}

\usepackage[caption=false,font=footnotesize]{subfig}

\usepackage{amsmath}
\usepackage{bbm}
\usepackage{stmaryrd}

\usepackage{array}
\usepackage{color}
\usepackage[hyphens]{url}
\usepackage[pdftitle=Title,pdfauthor=Anonymous]{hyperref}
\usepackage{xcolor}

\usepackage{listings}

\lstdefinestyle{interfaces}{
  float=t,
  floatplacement=t}

\usepackage{xspace}

\newcommand{\ie}{\textit{i.e.,}\xspace}
\newcommand{\eg}{\textit{e.g.,}\xspace}

\newenvironment{compactlistn}
  {\begin{enumerate} 
  \setlength{\itemsep}{0pt} 
  \setlength{\parskip}{0pt}} 
  {\end{enumerate}}

\newcommand{\dapp}{dapp\xspace}
\newcommand{\dapps}{dapps\xspace}

\IEEEoverridecommandlockouts
\IEEEpubid{\makebox[\columnwidth]{} \hspace{\columnsep}\makebox[\columnwidth]{ }}
\begin{document}

\pagenumbering{roman}

\title{Resolving the Multiple Withdrawal Attack on ERC20 Tokens}
\author{
	\IEEEauthorblockN{Reza Rahimian, Shayan Eskandari, Jeremy Clark}
	\IEEEauthorblockA{Concordia University}
}

\maketitle
\IEEEpubidadjcol

\begin{abstract}

Custom tokens are an integral component of decentralized applications (dapps) deployed on Ethereum and other blockchain platforms. For Ethereum, the ERC20 standard is a widely used token interface and is interoperable with many existing dapps, user interface platforms, and popular web applications (\eg exchange services). An ERC20 security issue, known as the \textit{multiple withdrawal attack}, was raised on GitHub and has been open since November 2016. The issue concerns ERC20's defined method  \texttt{approve()} which was envisioned as a way for token holders to give permission for other users and dapps to withdraw a capped number of tokens. The security issue arises when a token holder wants to adjust the amount of approved tokens from $N$ to $M$ (this could be an increase or decrease). If malicious, a user or dapp who is approved for $N$ tokens can front-run the adjustment transaction to first withdraw $N$ tokens, then allow the approval to be confirmed, and withdraw an additional $M$ tokens. In this paper, we evaluate 10 proposed mitigations for this issues and find that no solution is fully satisfactory. We then propose 2 new solutions that mitigate the attack, one of which fully fulfills constraints of the standard, and the second one shows a general limitation in addressing this issue from ERC20's approve method.

\end{abstract}


\IEEEpeerreviewmaketitle

\section{Introduction}

Ethereum is a public blockchain proposed in 2013, deployed in 2015~\cite{Ref00}, and has the second largest market cap at the time of writing\footnote{[2019-02-11] \url{https://coinmarketcap.com/currencies/ethereum/}}. It has a large development community which track enhancements and propose new ideas.\footnote{[2019-02-11] \url{https://www.coindesk.com/data}} Ethereum enables decentralized applications (\dapps) to be deployed and executed. \dapps can accept and transfer Ethereum's built-in currency (ETH) or might issue their own custom currency-like tokens for specific purposes. Tokens might be currencies with different properties than ETH. They may be required for access to a \dapp's functionality or they might represent ownership of some off-blockchain asset. It is beneficial to have interoperable tokens with other \dapps and off-blockchain webapps, such as exchange services that allow tokens to be traded.

\begin{figure}[t!]
	\centering
	\includegraphics[width=1.0\linewidth]{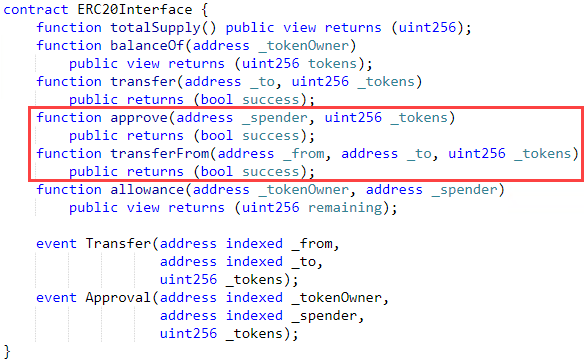}
	\caption{The ERC20 standard defines 6 methods and 2 events that must be implemented. Using \texttt{approve} and \texttt{transferFrom} in the existence of a race condition may lead to a \textit{``multiple withdrawal attack''}. Note that \texttt{transferFrom} augments the more basic \texttt{transfer} method.}\label{fig:erc20api}
\end{figure}

Toward this goal, the Ethereum project accepted a proposed standard (called ERC20~\cite{Ref08}) for a set of methods which ERC20-compliant tokens must implement. In terms of object oriented programming, ERC20 is an interface that defines abstract methods (name, parameters, return types) and provides guidelines on how the methods should be implemented, however it does not provide an actual concrete implementation (see Figure~\ref{fig:erc20api}). 

Since the introduction of ERC20 in November 2015, several vulnerabilities have been discovered. In November 2016, a security issue called \textit{``Multiple Withdrawal Attack''} was opened on GitHub~\cite{Ref13,Ref07}. The attack originates from two methods in the ERC20 standard for approving and transferring tokens. The use of these functions in an adverse environment (\eg front-running~\cite{eskandari2019sok}) could result in more tokens being spent than what was intended. This issue is still open and several solutions have been made to mitigate it. The authors of the ERC20 standard~\cite{Ref08} reference two sample implementations: OpenZeppelin~\cite{Ref10} and ConsenSys~\cite{Ref11}. OpenZeppelin mitigates the attack by introducing two additional methods to increase or decrease approved tokens (see Section~\ref{sec:mdao}), and the ConsenSys implementation does not attempt to resolve the attack. Additional implementations have a variety of different trade-offs in mitigating the issue (see Section~\ref{sec:eval}).

\begin{figure*}[t]
\includegraphics[width=\textwidth]{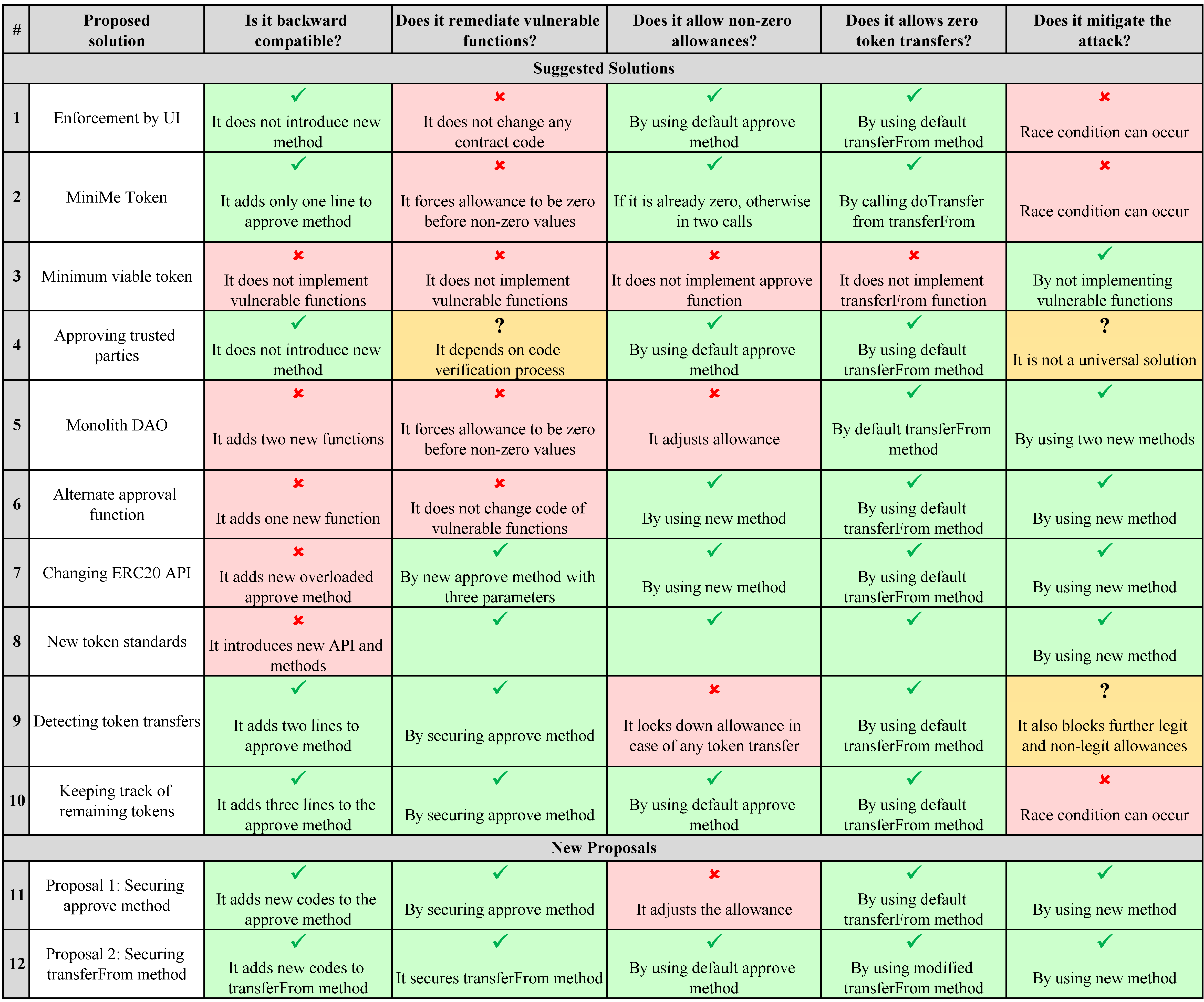}
\caption{Comparison of 10 proposals and 2 contributed by this paper. In our first proposal, CAS is used to mitigate the attack by comparing transferred tokens with new allowance. It is not fully ERC20-compliant since the allowance result does not always match what is requested. In our second proposal, a new local variable is defined to keep track of transferred tokens and prevents transfers in the case of already transferred tokens.\label{tab:comp}}
\end{figure*}

\subsubsection*{Contributions} In this paper, we evaluate 10 proposed mitigations for the \textit{``multiple withdrawal attack''}. We develop a set of criteria that encompass backwards compatibility, interoperability, adherence to the ERC20 standard, and attack mitigation. The summary is provided in Figure~\ref{tab:comp}. Since no mitigation is fully satisfactory, we develop two additional solutions based on the \textit{Compare and Set (CAS\footnote{A widely used lock-free synchronization strategy that allows comparing and setting values atomically.})} pattern\cite{Ref06}. We study in detail possible implementations of ERC20's \texttt{approve} and \texttt{transferFrom} methods. We argue that a CAS-based approach can never adequately deploy a secure \texttt{approve} method while adhering to the ERC20 standard. We then propose a secure implementation of the \texttt{transferFrom} method that mitigates the attack and fully satisfies the ERC20 standard. 


\section{Preliminaries}

\subsection{How the \textit{multiple withdrawal attack} works}
According to the ERC20 API definition, the \texttt{approve} function\footnote{We use the term method and function interchangeably.}
allows a spender (\eg user, wallet or other smart contracts) to withdraw up to an allowed amount of tokens from token pool of the approver. If this function is called again, it overwrites the current allowance with the new input value. On the other hand, the \texttt{transferFrom} function allows the spender to actually transfer tokens from the approver to anyone they choose (importantly: not necessarily themselves). The contract updates balance of transaction parties accordingly. 

An adversary can exploit the gap between the confirmation of the \texttt{approve} and \texttt{transferFrom} functions since the \texttt{approve} method replaces the current spender allowance with the new amount, regardless of whether the spender already transferred any tokens or not. This functionality of the \texttt{approve} method is shaped by the language of the standard and cannot be changed. Furthermore, while variables change and events are logged, this information is ambiguous and cannot fully distinguish between possible traces. Consider the following illustration:

\begin{compactlistn}
	\item Alice allows Bob to transfer N tokens on her behalf by broadcasting \texttt{approve(\_Bob, N)}.
	\item Later, Alice decides to change Bob's approval from N to M  by calling \texttt{approve(\_Bob, M)}.
	\item Bob notices Alice's second transaction after its broadcast to the Ethereum network but before adding to a block.
	\item Bob front-runs (using an asymmetric insertion attack~\cite{eskandari2019sok}) the original transaction with a call to \texttt{transferFrom(\_Alice, \_Bob, N)}. If a miner is incentivized (\eg by Bob offering high gas) to add this transaction before Alice's, it will transfer N of Alice's tokens to Bob.
	\item Alice's transaction will then be executed which changes Bob's approval to M.
	\item Bob can call \texttt{transferFrom} method again and transfer M additional tokens.
\end{compactlistn}

\begin{figure}[ht]
	\centering
	\includegraphics[width=1.0\linewidth]{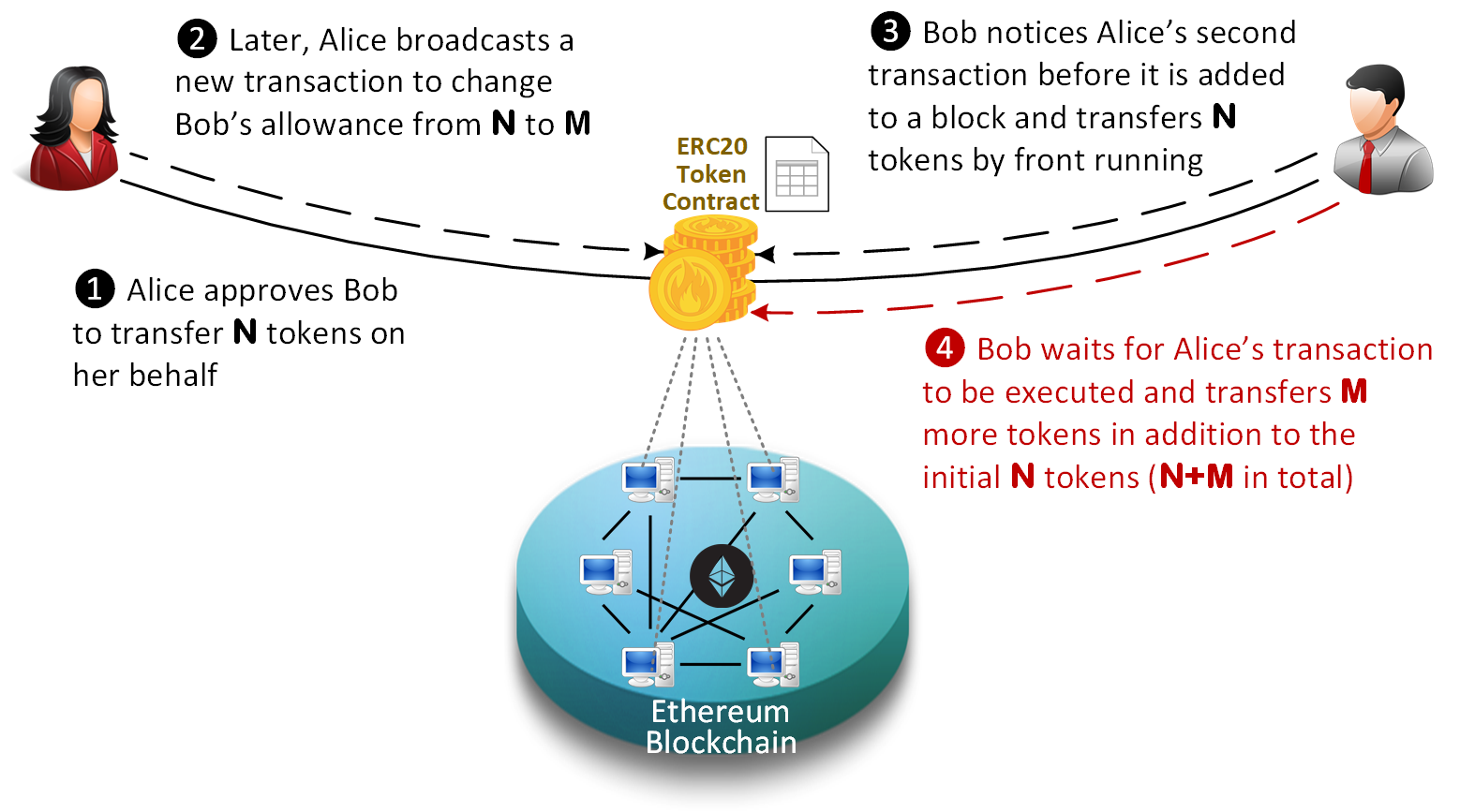}
	\caption{Possible \textit{``multiple withdrawal attack''} in ERC20 tokens when Alice changes Bob's allowance from N to M. By front-running, Bob is able to move total of N+M tokens from Alice. Possible mitigation should consider Bob's initial transfer of N tokens as legitimate transaction (step 3) and prevent the second transfer of M tokens (step 4).\label{fig:mwa}}
\end{figure}

In summary, in attempting to change Bob's allowance from N to M, Alice makes it possible for Bob to transfer N+M of her tokens. We operate on the assumption that a secure implementation would prevent Bob from withdrawing Alice's tokens multiple times when the allowance changes from N to M (see Figure~\ref{fig:mwa}).

\subsection{Why mitigation is important}
ERC20 tokens are important component of Ethereum's supplementary financial system that have many financial (as well as non-financial) uses and could hold considerable value (potentially exceeding the value of Ether itself). There has been more than 64,000 functional ERC20 tokens as of early 2019~\cite{victormeasuring} that might be vulnerable to this attack. Furthermore, ERC20 tokens that have already been issued cannot easily migrate to a new secure implementation and should these tokens appreciate in value in the future. Resolving the attack also serves as basis for other extended standards, such as ERC-777~\cite{Ref23} to be backward compatible with ERC20 interface~\cite{frowis2018detecting}. Finally, firms that hold ERC20 tokens require assurance of their security, particularly in the case that they require their financial statements to be audited---an issue like this could lead to further hesitation by auditors.

\subsection{Where to prevent this attack}

There are a few logical places to address this attack. Ideally the token author (instead of the token holders) would mitigate the attack within the ERC20 smart contract. Since two methods are involved in the attack, it could be addressed within the \texttt{approve} and/or \texttt{transferFrom} method. By contrast, token owners have no control over the implementation of the contract and are relegated to mitigate the attack by carefully monitoring the contract around the time allowance changes are made.

\subsubsection*{Prevention by token holder}\label{sec:preho}

Consider Alice, a token holder using a web app (\eg a user interface deployed with web3) to adjust Bob's allowance. If this user interface (UI) is written to the ERC20 standard, Alice will only have \texttt{approve} available to her. The ERC20 authors~\cite{Ref08} advise Alice against directly changing Bob's allowance from N to M. Instead, she should set the approval to 0 and then it to M (N$\rightarrow$0$\rightarrow$M). Presumedly, Alice will not do the second approval (setting it from 0 to M) if she sees that Bob withdraws N tokens before her N to 0 adjustment is confirmed. 

How will Alice know whether Bob withdraws first? The answer depends on how deeply her webapp can monitor the blockchain. One option is to monitor the variable that records Bob's allowance (technically an on-blockchain helper dapp could also do this). However if she sees Bob's allowance at 0 after initiating the first adjustment, it does not tell her that Bob did not withdraw N tokens---both methods result in Bob having an allowance of 0 so either (or both) could have been executed.

Next, she might rely on events passed from the contract to her web3 app. \texttt{Transfer} events as specified in ERC20 will log the transfer parameters (\ie  \texttt{address \_from, address \_to, uint256 \_tokens}). Alice's webapp could filter the events and only display transfers matching her address in the \texttt{\_from} field. The displayed transfers will include any transfer Bob makes, however Bob can provide any address in \texttt{\_to}, not just his own, and the event does not report who authorized the transaction (\ie \texttt{msg.sender}). If Alice has many authorizations, she cannot determine if a transfer was initiated by Bob or someone else she has authorized.\footnote{Even if Bob is listed in \texttt{\_to}, another authorized spender might have transferred tokens to Bob to make Alice believe Bob attempted a multiple withdrawal when he actually did not.} Therefore Alice cannot always unambiguously rely on events to determine Bob did not transfer funds (See Section~\ref{sec:enfui} for more details). If Alice's web app is beyond web3 and runs a full node maintaining blockchain state, it could correctly detect\footnote{By replaying transactions on an EVM} and attribute all transfers initiated by Bob. But this rather time-consuming and thus probably not efficient.

The takeaway from all these options is that prevention by token owners has some undesirable properties: (1) it splits adjustments into two transactions, (2) because the first transaction needs to be confirmed before the second is initiated, it takes time to complete, and (3) to precisely mitigate this attack, a heavyweight web app is needed to inspect deeper than variable state changes and events. For these reasons, we concentrate on mitigating the attack in the contract itself. If mitigation at the contract level works, allowances can be adjusted with a single function call from any existing lightweight ERC20 user interface, and no additional monitoring of the contract is necessary. 

\subsubsection*{Prevention by token author in \texttt{approve}} The next logical place to tackle multiple withdrawal is in the implementation of the \texttt{approve} method. In particular, an \texttt{approve} implementation might be engineered to fail under the conditions of multiple withdrawal, to adjust the approval amount, treat adjustments as relative offsets from the current amount, or other techniques. As we review the 10 solutions, we will see different proposals along these lines, as well as our own proposal in Section~\ref{sec:proposal1}. For now, we emphasize that adherence to the standard is a core challenge as it unequivocally states that: ``If this function is called again, it overwrites the current allowance with \texttt{\_value}''~\cite{Ref08}. So, any adjustment violates the standard.

\subsubsection*{Prevention by token author in \texttt{transferFrom}} Recall from Figure~\ref{fig:mwa} that step 4 is the offending function call and it is \texttt{transferFrom}. If we add new state to the contract to track the number of tokens that have been transferred, we can allow approval to work exactly as specified while interpreting it as a ``lifetime'' allowance. We will explain this in more detail in Section~\ref{sec:proposal2}. 

\subsection{What an ideal solution looks like}\label{sec:criteria}
We prioritize adherence to the ERC20 standard. While deviating from the standard might become acceptable if there is no possible way to conform with it and maintain security, we consider that a last resort. Indeed, as we will show, it is possible to secure an ERC20 contract within the constraints of the standard~\cite{Ref08}, which we summarize here:

\begin{compactlistn}
\item The input to \texttt{approve} method is a new allowance and not a relative adjustment.
\item The result of \texttt{approve} method will overwrite the current allowance with the new allowance.
\item A call to \texttt{transferFrom} on an input of 0 tokens will execute as a normal transfer and emit a \texttt{Transfer} event.
\item A spender can call \texttt{transferFrom} multiple times up to the allowed amount.
\item Transferring up to any initial allowance is always a legitimate transfer.
\item An ideal solution cannot rely on overloading existing methods or introducing new methods outside of ERC20, as existing ERC20 dapps and web apps would have to be modified to interoperate.
\item A solution must eliminate all race conditions.
\end{compactlistn}


\section{Evaluating Proposed Solutions}\label{sec:eval}

In this section, we evaluate 10 solutions that have been proposed by Ethereum community\footnote{Mostly from developers on GitHub} to address the multiple withdrawal attack. We examine each solution in detail and evaluate them against the criteria established in section~\ref{sec:criteria}. The summary is also presented in Figure~\ref{tab:comp}.


\subsection{Enforcement by User Interface (UI)}\label{sec:enfui}

\begin{figure}[t!]
	\centering
	\includegraphics[width=1.0\linewidth]{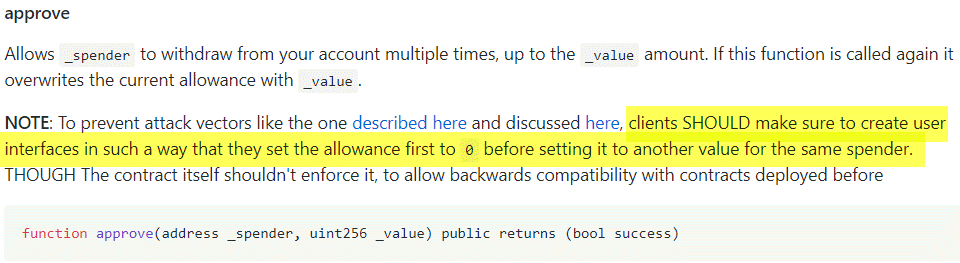}
	\caption{Recommendation of ERC20 standard to mitigate \textit{``multiple withdrawal attack''} by enforcement in UI.\label{fig:uie}}
\end{figure}

The first solution is enforcement at the user interface level. We discussed this previously in section~\ref{sec:preho} but we reiterate the main points here again. The exact recommendation from the ERC20 standard is shown in Figure~\ref{fig:uie} and is essentially to set an allowance to zero before any non-zero values. Presumedly, it will also enforce that the new approval is not allowed if proceeded by a token transfer by the approved spender. We consider the UI to be a lightweight web app that can reference a contract's state variables and emit events but does not maintain a full copy of the blockchain. Consider (again) the most basic attack sequence:

\begin{enumerate}
	\item Alice allows Bob to transfer N of her tokens.
	\item Alice's client broadcasts an allowance of 0.
	\item Bob broadcasts a competing transaction to transfer N of Alice’s tokens.
	\item Bob's transaction front-runs Alice's, is confirmed, and sets Bob’s allowance to 0 (from N).
	\item Alice’s transaction is confirmed and sets Bob’s allowance to 0 (from 0).
	\item Alice's client broadcasts an allowance of M.
	\item Alice’s second transaction is confirmed and sets Bob's allowance to M.
	\item Bob transfers M of Alice’s tokens for a total of N+M tokens.
\end{enumerate}

The key mitigation to this attack is for Alice's client to pause at step 6 and determine if a transaction sequence like 3\&4 has occurred or not. This cannot be determined by monitoring the integer that records Bob's allowance because it will be 0 regardless of whether steps 3\&4 occurred or not.

It also cannot always be determined by monitoring the events emitted from the contract. Step 4 will log a transfer from Alice's address to Bob's address of N tokens. If no event is emitted, Alice can know for certain no transfer was made. However if an event is emitted, Alice must decide it was a transfer initiated by Bob or a transfer by someone else she has authorized. If she has not authorized anyone else, she can know for certain it was Bob. However if she has a busy account with multiple authorized spenders of her tokens, the event is not verbose enough to determine what happened. Importantly, it does not record who (\texttt{msg.sender}) initiated the transfer, only who received the tokens, and these are not necessarily the same entity. Bob can send Alice's tokens to his accomplice Eve, or some other authorized spender can send Alice's tokens to Bob (which looks like Bob is attacking when he is not). Since a UI is automated and does not use human discretion, it cannot decide circumstantially whether something looks like an attack or not---it must either specify exact rules, which it cannot do here because of the ambiguity of the events, or it can ask for Alice's human input which introduces usability issues. In conclusion, it is better for enforcement to happen at the contract level as implemented by the most of the solutions.


\subsection{MiniMeToken}\label{sec:MiniMeToken}

MiniMeToken~\cite{Ref15} enforces the recommendation that allowances are first set to zero before setting to a non-zero value. The enforcement is done within the ERC20 implementation by adding \texttt{require((\_amount == 0) || (allowed[msg.sender][\_spender] == 0))} to the \texttt{approve} method. (\texttt{\_amount == 0}) allows approvals to be set to 0, and the second condition requires the allowance of \texttt{\_spender} to be 0 before it can be set to a non-zero value. This solution fails to prevent Bob from transferring N+M tokens, as the contract will be unable to determine whether N tokens have been already drained by Bob or not. Recall the attack sequence specified in section~\ref{sec:enfui}. In step 5, Alice finds Bob's allowance is set to zero. However, she cannot distinguish whether it was because of her transaction or token transfer by Bob\footnote{Both \texttt{approve} and \texttt{transferFrom} can set Bob's allowance to zero. In the first case, if Alice's transaction executes before Bob's and in the second one due to token transfer by Bob.}. Furthermore, contracts cannot read their own logs to filter out \texttt{Transfer} events.


\begin{figure}[t]
	\centering
	\includegraphics[width=1.0\linewidth]{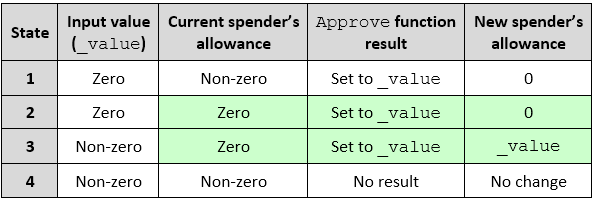}
	\caption{Functionality of the \texttt{approve} method in MiniMeToken and MonolithDAO tokens, which are implemented slightly differently but effectively enforce setting spender's allowance in two steps: first to zero and then to any non-zero value (\eg N$\rightarrow$0$\rightarrow$M).\label{fig:dao}}
\end{figure}

\subsection{MonolithDAO}\label{sec:mdao}
MonolithDAO Token~\cite{Ref12} (and its extension in OpenZeppelin~\cite{Ref10}) implement two additional functions for approval increases and decreases: \texttt{increaseApproval} and \texttt{decreaseApproval}. Both take two parameters: the address of the approved spender and the amount to be added/subtracted from the spender's current approval. Additionally, the \texttt{approve} method has additional logic to enforce that owners set the allowance to zero before non-zero values (see Figure~\ref{fig:dao}). Forcing a set to zero is enforced the same way as in MiniMeToken~\ref{sec:MiniMeToken}. After the initial approval, owners use \texttt{decreaseApproval} and \texttt{increaseApproval} instead of \texttt{approve}. Consider the following transaction sequence:

\begin{enumerate}
	\item Alice allows Bob to transfer N tokens by broadcasting \texttt{approve(\_Bob, N)}\footnote{Alice can use the default \texttt{approve} method since Bob’s allowance was initially 0.}. 
	\item Alice increases Bob’s allowance to M. Since execution of \texttt{approve(\_Bob, M)} will fail (Bob's allowance is non-zero), Alice broadcasts \texttt{increaseApproval(\_Bob, M-N)} instead.
	\item Bob front-runs this transaction by broadcasting \texttt{transferFrom(\_Alice, \_Bob, N)}.
	\item If Bob's transaction is confirmed first, he will transfer N tokens but this is legitimate as he was approved for N, and his approval is set to 0.
	\item Alice's transaction will adjust Bob's approval from 0 to M-N. In total, Bob would be able to spend N+M-N=M tokens as intended.
	\item Remark: for decreases, \texttt{decreaseApproval} will similarly prevent the attack by reducing Bob's allowance instead of setting it to a particular value.
\end{enumerate}

Although these two new complementary functions do prevent the attack, they have not been defined in the initial specifications of ERC20 standard. Therefore, these safer functions will not be used by ERC20-compliant web apps and smart contracts that are already deployed. Such deployments will continue to use \texttt{approve} method which suffers the same issue as MiniMeToken: set to zero does not always mitigate the attack. 


\subsection{Detecting token transfers}

\begin{figure}[t]
	\centering
	\includegraphics[width=1.0\linewidth]{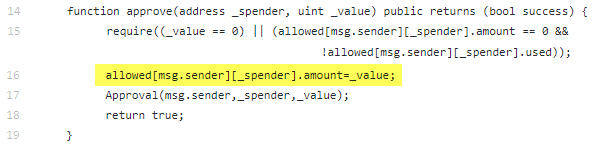}
	\caption{In \textit{``Detecting token transfers''}, the \texttt{approve} method needs to be modified by adding a line of code like \texttt{allowed[msg.sender][\_spender].used = false;} between lines 16 and 17 to unlock spender flag for the next legitimate change. However, this change makes attack mitigation ineffective.\label{fig:det}}
\end{figure}

The next approach \cite{Ref17} maintains a small state machine to enforce prevent approvals preceded by transfers. We revisit and elaborate on this approach in our own solution in Section~\ref{sec:proposal1}. In this proposal, the implementation of \texttt{approve} is augmented with a flag (one flag for each pair of owners and approved spenders) that can be set in \texttt{transferFrom} function. \texttt{transferFrom} sets the flag to \texttt{true} after any token transfer and the \texttt{approve} method requires the flag to be \texttt{false} before allowing new approvals. This approach requires a new data structure, can prevent front-running, but also has a deadlock scenario. Consider the following scenario:

\begin{enumerate}
	\item Alice allows Bob to transfer N tokens by broadcasting \texttt{approve(\_Bob, N)}. This succeeds because Bob’s allowance was initially 0 and his corresponding flag=\texttt{false} (line 15 in Figure~\ref{fig:det}).
	\item Bob legitimately withdraws N tokens by calling \texttt{transferFrom(\_Alice, \_Bob, N)}.
	\item Alice decides to increase Bob's allowance to M by broadcasting \texttt{approve(\_Bob, 0)}.
	\item Since Bob's transaction is confirmed first, \texttt{transferFrom} turns his flag to \texttt{true}.
	\item Alice’s transaction is confirmed, passes the check because input value is 0 (line 15), and Bob’s allowance is set to 0 while his flag remains \texttt{true}. Critically, the \texttt{approve} method does not flip the spender's flag.
	\item Alice wants to change Bob’s allowance to M and broadcasts \texttt{approve(\_Bob, M)}. 
	\item Since Bob already transferred N tokens (or it could be just 1 token) and his flag=\texttt{true}, the transaction fails.
	\item Remark: Bob’s allowance is 0 and it cannot not change, so he is locked out of further approvals.
\end{enumerate}

Although this approach mitigates the attack, it prevents any further legitimate approvals. As discussed, in case of any legitimate token transfer by Bob, Alice would not be able to change his approval. Because Bob's flag is set to \texttt{true} and line 15 in Figure~\ref{fig:det} does not allow changing the allowance (an exception is thrown). A potential bypass is to set the allowance to 0 and then to M. However this transaction sequence never flips the flag to \texttt{false} (there is no code for it in the \texttt{approve} method). So it keeps Bob locked out of any further legitimate allowances. A quick fix might be having \texttt{approve} set the flag to \texttt{false} (\ie between lines 16 and 17). But this will cause another problem: after setting the allowance to 0, the spender flag becomes \texttt{false}, and allows non-zero values even if tokens have been already transferred. This will no longer prevent multiple withdrawals as demonstrated in the following transaction sequence:

\begin{enumerate}
	\item Alice changes Bob's allowance from N to 0.
	\item Bob transfers N tokens before the allowance change and his \texttt{used} flag turns to \texttt{true}.
	\item Alice's transaction is successful since input \texttt{\_value} is 0. The second condition is not evaluated although \texttt{used} flag is \texttt{true}.
	\item Alice's transaction turns \texttt{used} flag to \texttt{false} and sets Bob's allowance to 0. 
	\item Now Alice wants to set Bob's allowance from 0 to M, his flag is \texttt{false} and allowance is 0. 
	\item Remark: Alice cannot distinguish whether Bob moved any token or not. Setting a new allowance will allow Bob to transfer more tokens.
\end{enumerate}

In fact, resetting the flag in \texttt{approve} method will not fix the issue and makes attack mitigation ineffective. In short, this approach can not satisfy both legitimate and non-legitimate scenarios. Nevertheless, it is a step forward by introducing the need for a new state to track transferred tokens if a solution is to be found.


\subsection{Keeping track of remaining tokens}

\begin{figure}[t]
	\centering
	\includegraphics[width=1.0\linewidth]{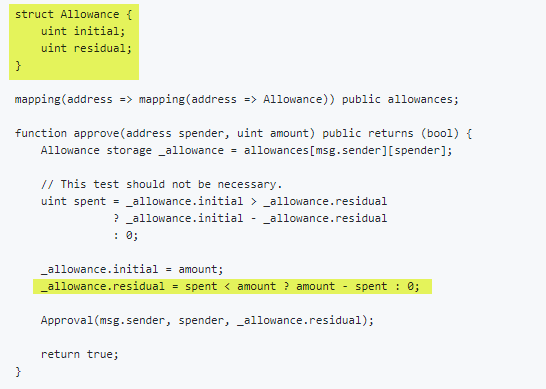}
	\caption{Keeping track of remaining tokens by introducing a new data structure.\label{fig:track}}
\end{figure}

This approach~\cite{Ref18} is inspired by the previous solution of detecting token transfers and introduces new state in the contract. It keeps track of the remaining tokens, and an ERC20-compliant \texttt{approve} method uses these variables to set allowances accordingly (see Figure~\ref{fig:track}).

At a first glance, it seems to be a promising solution by more effectively enforcing approvals to zero before non-zero values. However, the highlighted code in \texttt{approve} method (see Figure~\ref{fig:track}) resembles the situation that is explained in~\ref{sec:enfui}. In case of front-running, both \texttt{initial} and \texttt{residual} variables will be zero and it would not be possible for Alice to distinguish if any token transfer have occurred due to her allowance change or due to Bob transferring a token. To illustrate this, considering the following transaction sequence (For formatting reasons, we abbreviate codes like \texttt{allowances[\_Alice][\_Bob].initial} as \texttt{AB.initial}):

\begin{enumerate}
	\item Bob’s allowance is initially zero (\texttt{AB.initial=0}) and his residual is zero as well (\texttt{AB.residual=0}).
	\item Alice allows Bob to transfer N tokens and makes \texttt{AB.initial=N} and \texttt{AB.residual=N}.
	\item Alice decides to change Bob’s allowance to M and has to set it to zero before M.
	\item Bob notices Alice’s broadcast and front-runs it with a transfer of N tokens.
	\item Consequently, the \texttt{transferFrom} function sets his residual to zero (\texttt{AB.residual=0}).
	\item Alice’s transaction for setting Bob's allowance to 0 is confirmed and sets \texttt{AB.initial=0} and \texttt{AB.residual=0}.
	\item Remark: at this stage, the state is indistinguishable from Step 1. Alice cannot distinguish whether any token have been transferred or not based on \texttt{AB.initial} and \texttt{AB.residual}. 
	\item Alice approves Bob for spending new M tokens and Bob is able to transfer new M tokes in addition to initial N tokens.
\end{enumerate}

It is true that a \texttt{Transfer} event has been recorded as a result of step 5. However transfer events are ambiguous as described in Section~\ref{sec:enfui}. Thus it is not always possible for the approver to detect legitimate from non-legitimate tokens transfers. Overall, this approach cannot prevent the attack in the case of front-running by Bob. 


\subsection{Overloading \texttt{Approve} method}
\label{sec:overload}

\begin{figure}[t]
	\centering
	\includegraphics[width=1.0\linewidth]{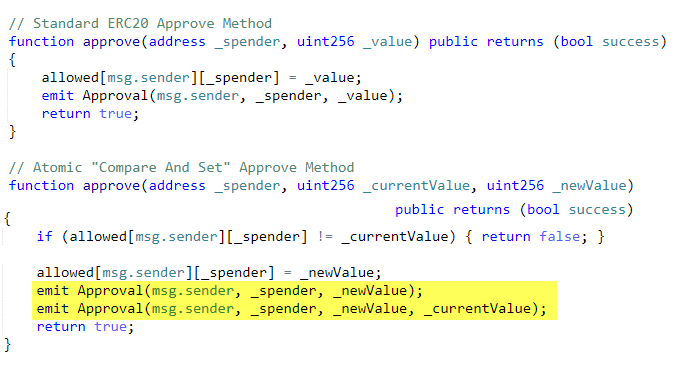}
	\caption{An overloaded approve method adds a method with three parameters to compare and set new allowance atomically.\label{fig:api}}
\end{figure}

As advised by \cite{Ref03}, a secure \texttt{approve} method could take one additional parameter: the expected allowance of the spender when the adjustment to the approval is made. Under this proposal, the adjustment succeeds only if the passed expected allowance matches the spender's actual allowance, and fails otherwise. Consider a multiple withdrawal attack where Bob is approved for N of Alice's tokens, Alice adjusts his approval from N to M tokens, and Bob front-runs the approval with a transfer of N tokens. Alice's approval will specify N as the current expected allowance when adjusting it to M. Because Bob's transfer of N tokens changes his approval to 0, Alice's approval will fail because it expects an allowance of N when the allowance is 0. This allows atomic compare and set of the spender's allowance to make the attack impossible. 

While this approach mitigates the attack, it requires a new overloaded \texttt{approve} method with three parameters, in addition to the standard ERC20 \texttt{approve} method with two parameters (see Figure~\ref{fig:api}). Additionally it defines a new event. Existing ERC20 web apps and smart contracts will be unaware of the overloaded method and continue to call the insecure two parameter method. Thus it does not provide backward compatibility and interoperability with already deployed smart contracts. 


\subsection{Alternate approval function}

Another suggestion \cite{Ref16} is to move the security check to a new function called \texttt{safeApprove}\footnote{Syntax: \texttt{safeApprove(address \_spender, uint256 \_currentValue, uint256 \_value}} that compares the current and new allowance values (like the overloaded \texttt{approve} in Section~\ref{sec:overload}). The adjustment is only allowed if the allowance has not been changed by the time the function is executed. In this case, Alice uses the standard \texttt{approve} function to set Bob’s allowance to 0 and for new approvals, she has to use \texttt{safeApprove} function. As above, \texttt{safeApprove} takes the current expected approval amount as input parameter and calls \texttt{approve} method if previous allowance is equal to the current expected approval. Although this approach mitigates the attack by using the CAS pattern~\cite{Ref06}, it is not interoperable with ERC20-compliant web apps and smart contracts that will be unaware of \texttt{safeApprove}.


\subsection{Minimum viable token}

Rather than adding new methods to ERC20, methods can be also taken away. We can reduce the ERC20 standard to a set of core functionalities and implement only the essential methods. The attack can be side-stepped if methods like \texttt{approve} and \texttt{transferFrom} are simply not implemented (recall that \texttt{transferFrom} is in addition to the more commonly used \texttt{transfer}) or are implemented to always throw an exception and revert. Golem Network Token (GNT\footnote{https://etherscan.io/address/0xa74476443119A942dE498590Fe1f2454d7\newline D4aC0d\#code}) is one of these examples since it does not implement the \texttt{approve}, \texttt{allowance} and \texttt{transferFrom} functions. According to the ERC20 specification~\cite{Ref08}, these methods are not \textsc{optional} and must be implemented. Moreover, ignoring them can cause failed function calls from smart contracts or web apps that expect these methods to work as specified. Therefore, we categorize this approach as successfully mitigating the attack but not offering interoperability.


\subsection{New token standards}

\begin{table*}
\centering
\def\arraystretch{1.2}%
\begin{tabular}{|m{1.8cm}|m{14.5cm}|}
	\hline\centering
	\textbf{Token Standard} & \textbf{A description of non-compliance with ERC20}\\
	\hline\hline\centering
	ERC 223 \cite{Ref20} & It does not implement ERC20 \texttt{approve} and \texttt{transferFrom} methods by assuming that they are potentially insecure and inefficient.\\ 
	\hline\centering 
	ERC 667 \cite{Ref21} & It solves the problem of the transfer function in ERC223 (\ie the need to implement \texttt{onTokenTransfer} routing in the receiving contract). It mitigates the attack using the same code as ERC223 with a supplementary function.\\ 
	\hline\centering 
	ERC 721 \cite{Ref22} & Unlike ERC20 tokens that share the same characteristics, ERC721 tokens are non-fungible tokens (NFT) where each token is unique and not interchangeable. In addition to this functional difference, ERC721does not implement \texttt{transferFrom} method of ERC20 standard and introduces a safe transfer function called \texttt{safeTransferFrom}.\\ 
	\hline\centering
	ERC 777 \cite{Ref23} & It does not implement \texttt{transfer} or \texttt{transferFrom} methods and replaces them with safe \texttt{send} and \texttt{operatorSend} methods. Moreover, It considers costly \texttt{approve}/\texttt{transferFrom} paradigm to be replaced by \texttt{tokensReceived} function. Therefore, ERC777 would not be backward compatible by this replacement. A token might implement both ERC20 and ERC777 but the ERC20 methods would require attack mitigation.\\ 
	\hline\centering 
	ERC 827 \cite{Ref24} & It uses OpenZeppelin's~\cite{Ref10} ERC20 implementation and defines three new functions to allow users for transferring data in addition to value in ERC20 transactions. This feature enables ERC20 tokens to have the same functionality as Ether (transferring data and value). In fact, it extends functionality of ERC20 tokens and not addressing the attack. \\ 
	\hline\centering 
	ERC 1155 \cite{Ref25} & It is improved version of ERC721 by allowing each Token ID to represent a new configurable token type, which may have its own metadata, supply and other attributes. ERC1155 aimed to remove the need to "approve" individual token contracts separately. Therefore, it does not implement any code to address this vulnerability.\\ 
	\hline\centering 
	ERC 1377 \cite{Ref26} & It implements \texttt{approve} method with three parameters in addition to the ERC20 default \texttt{approve} with two inputs. Additionally, it uses OpenZeppelin~\cite{Ref10} approach for increasing and decreasing approvals. We would consider it as mix of MiniToken and OpenZeppelin proposals that we discussed before.\\
	\hline
\end{tabular}
\newline
\caption{Evaluation of standard's adherence to ERC20 and mitigation of the multiple withdrawal attack.\label{tab:erc}}
\end{table*}

Minimum viable tokens could alternatively be considered a new non-ERC20 token. In fact, there are many alternatives that extend or modify ERC20 for a variety of purposes, mostly around functionality but some address multiple withdrawals. We summarize the main proposals in Table~\ref{tab:erc}. Despite the enhancements of these new token standards for future deployments, ERC20 is ingrained in the community and industry with 168,092 deployed tokens\footnote{https://etherscan.io/tokens, Accessed 18-Feb-2019}, many interoperable developer tools and libraries, and web platforms built on trading these tokens. Ideally, and the goal of this paper, a backward compatible solution could be found that does not change the ERC20 API or require token migration to a new standard (which is not necessarily possible to do at the contract level). Like minimum viable tokens, we categorize these approaches potentially mitigating the attack (depending on which standard --- see Table~\ref{tab:erc}) but not offering interoperability.


\subsection{Approving trusted parties}

A final solution is to limit token transfer approvals to trusted entities. Such a solution is discretionary---it cannot be automated within a contract---so it adds additional burden for the user. At first glance, it seem that Alice would never authorize Bob to spend her tokens if she does not trust Bob. However approvals are constrained to specific amounts specifically to enable some less trustworthy interactions. The \textit{``multiple withdrawal attack''} is damaging because Bob can circumvent the constraints. If Bob is another smart contract, instead of a user, then Alice could confirm it does not have the logic to conduct a multiple withdrawal attack, cannot be updated (\eg does not delegate function calls to code at other addresses), and thus it can be trusted with insecure ERC20 tokens. This is a sensible approach but it is quite limited to specific approval scenarios.


\section{New mitigations}

By this point, we have discussed 10 solutions to the multiple withdrawal attack and we evaluated them in terms of compatibility with the standard and attack mitigation (recall the summary in Figure~\ref{tab:comp}). Since none of them precisely satisfy the constraints of ERC20 standard, we now propose two new solutions to mitigate the attack.


\subsection{Proposal 1: Securing \texttt{approve} method}
\label{sec:proposal1}

\begin{figure}[t]
	\centering
	\includegraphics[width=1.0\linewidth]{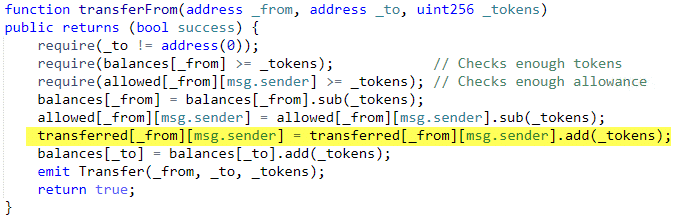}
	\caption{Modified version of \texttt{transferFrom} for keeping track of transferred tokens per spender.\label{fig:transfer1}}
\end{figure}

\begin{figure}[t]
	\centering
	\includegraphics[width=1.0\linewidth]{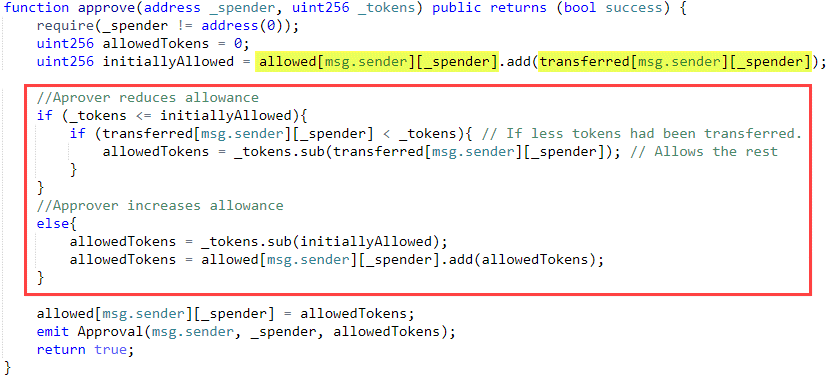}
	\caption{Added code block to \texttt{approve} function to prevent the attack by comparing and setting new allowance atomically.\label{fig:approve1}}
\end{figure}

By implementing the CAS pattern~\cite{Ref06} in the \texttt{approve} method, we set up a small state machine so that new allowances can be set atomically after a comparison with transferred tokens. This tracking requires adding a new variable to the \texttt{transferFrom} method (see Figure~\ref{fig:transfer1}). Since this is an internal variable, it is not visible to already deployed smart contracts and keeps the \texttt{transferFrom} function ERC20-compatible. Similarly, a block of code is added to the \texttt{approve} function (see Figure~\ref{fig:approve1}) to work in both cases with zero and non-zero allowances. This new logic in the \texttt{approve} function compares a new allowance---passed as \texttt{\_tokens} argument to the function---with the current allowance of the spender and the already transferred tokens. Allowance are saved in \texttt{allowed[msg.sender][\_spender]} variable as in typical ERC20 implementation, and \texttt{transferred[msg.sender][\_spender]} is the new state. The method decides to increase or decrease the current allowance based on this comparison. If the new allowance is less than initial allowance---sum of \texttt{allowance} and \texttt{transferred} variables---it denotes decreasing of allowance, otherwise increasing of allowance is intended. Such a modified \texttt{approve} function prevents the attack by either increasing or decreasing the allowance instead of setting it to an explicit value.

Unlike other solutions, there is no need to set allowance from N to 0 and then to M. The token holder can directly change the allowance from N to M which saves time waiting for the confirmation of a transaction and any monitoring of the contract. Consider the following transaction sequences to illustrate how the state changes:


\subsubsection*{Scenario A} Alice approves Bob for spending 100 tokens and then decides to increase it to 120 tokens.
\begin{enumerate}
	\item Alice approves Bob for transferring 100 tokens.
	\item After a while, Alice decides to increase Bob’s allowance from 100 to 120 tokens.
	\item Bob noticed Alice’s new transaction and transfers 100 tokens by front-running.
	\item Bob’s allowance is 0 and \texttt{transferred}=100.
	\item Alice’s transaction is mined and checks initial allowance (100) with new allowance (120).
	\item As it is increasing, the new allowance (120) will be subtracted from the transferred tokens (100).
	\item 20 tokens will be set as Bob’s allowance.
	\item Bob would be able to transfer 20 more tokens (120 in total as Alice wanted).\newline
\end{enumerate}
 
\subsubsection*{Scenario B} Alice approves Bob for spending 100 tokens and then decides to decrease it to 10 tokens.
\begin{enumerate}
	\item Alice approves Bob for transferring 100 tokens.
	\item After a while, Alice decides to reduce Bob’s allowance from 100 to 10 tokens.
	\item Bob noticed Alice’s new transaction and transfers 100 tokens by front-running.
	\item Bob’s allowance is 0 and \texttt{transferred}=100 (set by \texttt{transferFrom} function).
	\item Alice’s transaction is mined and checks initial allowance (100) with new allowance (10).
	\item As it is reducing, \texttt{transferred} tokens (100) is compared with new allowance (10). Since Bob already transferred more tokens, his allowance will be set to 0.
	\item Bob is not able to move more than initial 100 approved tokens.
\end{enumerate}


\begin{figure}[t]
	\centering
	\includegraphics[width=1.0\linewidth]{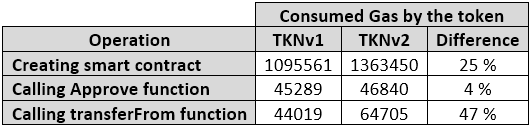}
	\caption{Comparison of gas consumption between standard implementation of ERC20 token (TKNv1) and secured version of it (TKNv2).\label{fig:gas}}
\end{figure}

\subsubsection*{Performance} In order to evaluate functionality of the new \texttt{approve} and \texttt{transferFrom} functions, we have implemented a standard ERC20 token (TKNv1\footnote{https://rinkeby.etherscan.io/address/0x8825bac68a3f6939c296a40f c8078d18c2f66ac7}) along side the proposed ERC20 token (TKNv2\footnote{https://rinkeby.etherscan.io/address/0xf2b34125223ee54dff48f715 67d4b2a4a0c9858b}) on the Rinkeby test network. Our testing for different input values shows that TKNv2 can address \textit{``multiple withdrawal attack''} by making front-running gains ineffective. Moreover, we compared these two tokens in term of gas consumption. TKNv2.\texttt{approve} function uses almost the same amount of gas as TKNv1.\texttt{approve}, however, gas consumption of TKNv2.\texttt{transferFrom} is around 47\% more than TKNv1.\texttt{transferFrom} (see Figure~\ref{fig:gas}). This difference in TKNv2 is because of maintaining a new mapping variable for tracking transferred tokens. In term of compatibility, both are equivalent interoperable with standard wallets (\eg MetaMask and MEV) and have not raised any transfer issues.

\subsubsection*{Discussion} In summary, we can use the CAS pattern to implement a secure \texttt{approve} method that can mitigate the attack effectively. However, it violates one of the ERC20 specifications that says: ``If \texttt{approve} function is called again, it overwrites the current allowance with \texttt{\_value}" (item 2 in Section~\ref{sec:criteria}). Our solution does not comply with this as the resulting allowance can be different than what is passed by the approver (as shown in the scenarios above). Furthermore we argue that is in fact impossible to secure the \texttt{approve} method without adjusting the allowance. Considering the following transaction sequence:

\begin{enumerate}
	\item Alice decides to change Bob's allowance from N to M (M $\leq$ N in this example).
	\item Bob transfers N tokens by front-running and the \texttt{transferred} variable sets to N.
	\item Alice's transaction is mined and the \texttt{approve} method detects Bob's token transfer.
	\item If \texttt{approve} method does not adjust the allowance based on transferred tokens, it has to set it to M---to conform with the standard---which is allowing Bob to transfer more M tokens, or it could fail which deadlocks Bob from future approvals\footnote{Consider Alice wants to allow Bob for transferring M more tokens in addition to initial N tokens (N+M tokens in total). So, she passes M+N to the \texttt{Approve} method. Even in case of front-running by Bob, the \texttt{Approve} method should not throw an exception. Because this is a legitimate withdraw and already approved by Alice. Additionally, there would not be a way of detecting front-running in \texttt{Approve} method. It sees only transferred token without knowledge of their previous value.}.
\end{enumerate}

Therefore the \texttt{approve} method has to adjust the allowance according to transferred tokens, not based on passed input values to the \texttt{approve} method. Overall, there seems to be no solution to secure the \texttt{approve} method while adhering specification of ERC20 standard.


\subsection{Proposal 2: Securing \texttt{transferFrom}}\label{sec:proposal2}

\begin{figure}[t]
	\centering
	\includegraphics[width=1.0\linewidth]{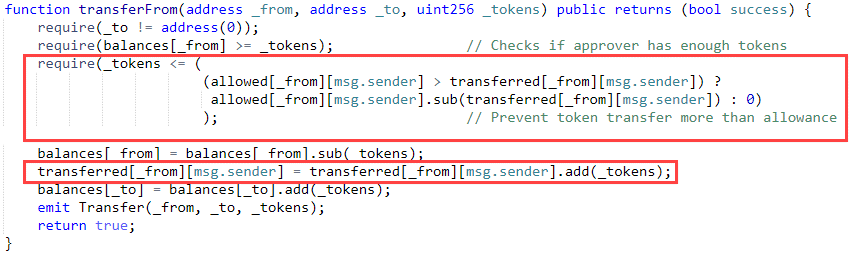}
	\caption{Securing \texttt{transferFrom} method instead of \texttt{approve} method can mitigate the attack by preventing more token transfer than allowed.\label{fig:transfer2}}
\end{figure}

\begin{figure}[t]
	\centering
	\includegraphics[width=1.0\linewidth]{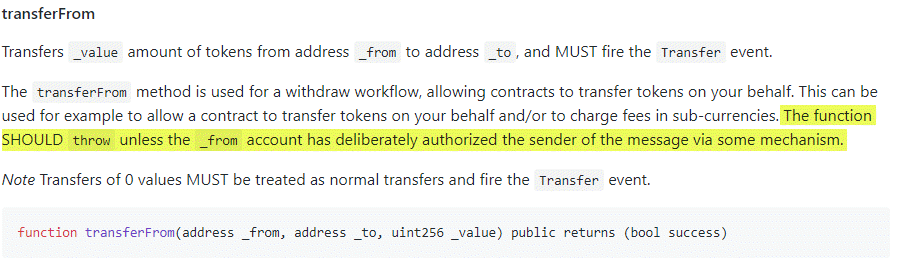}
	\caption{ERC20 \texttt{transferFrom} method definition that emphasizes on throwing an exception when the spender is not authorized to move tokens.\label{fig:standard}}
\end{figure}

As an alternative to Proposal 1, we can also consider securing the \texttt{transferFrom} method. As specified by the ERC20 standard (see figure~\ref{fig:standard}), the goal here is to prevent the spender from transferring more tokens than allowed. Based on this assumption, we should not rely solely on the allowance value in deciding whether to allow or prevent an approve and should also consider the number of transferred tokens, which requires new state as in Proposal 1. 

Our solution, which is compliant with a careful reading of ERC20, is to interpret allowance as a `global' or `lifetime' allowance value, instead of the amount allowed at the specific time of invocation. For example, say Alice approves Bob for 50 tokens, Bob transfers 50 tokens, Alice approves Bob for 30 (more) tokens, and Bob transfers 30 tokens. In our implementation, Alice would approve Bob for 50 tokens and he transfers 50 tokens. To approve Bob for 30 more tokens, she approves Bob for 80 tokens. He has already spent 50 of these 80 tokens so he will only be allowed to transfer an addition 30. Thus 80 is his lifetime allowance and 50 (kept internally) is the amount he has transferred. In a bit more detail, consider the following, which prevents multiple withdrawals by modifying the implementation of \texttt{transferFrom} but keeping \texttt{approve} untouched:

\begin{enumerate}
	\item Alice approves Bob to transfer 100 tokens
	\item Alice broadcasts an approval of 70, decreasing Bob's allowance.
	\item Bob front-runs Alice’s transaction and transfers 100 tokens (remark: a legitimate transfer).
	\item Alice's transaction is confirmed and sets Bob allowance to 70 by the default \texttt{approve} method.
	\item Bob's noticed the new allowance and tries to move 70 additional tokens by broadcasting \texttt{transferFrom(\_Bob,70)}. 
	\item Since Bob has already transferred more than 70 tokens, his transaction fails and prevents multiple withdrawal. 
	\item In the end, Bob’s allowance is set at 70 and his transferred tokens are set at 100.
\end{enumerate}

\subsubsection*{Performance \& Discussion} Interpreting \texttt{allowance} as a lifetime allowance is completely in accordance with the ERC20 standard (see figure~\ref{fig:standard}). In our solution, there is no relation between allowance (\texttt{allowed[\_from][msg.sender]}) and transferred tokens (\texttt{transferred[\_from][msg.sender]}). The first variable shows lifetime transferable tokens by a spender and can be changed independently of the transferred tokens (\ie \texttt{approve} method does not check transferred tokens). If Bob has not already transferred that many tokens, he would be able to transfer the difference of it. Our token is implemented as TKNv3\footnote{https://rinkeby.etherscan.io/address/0x5d148c948c01e1a61e280c8 b2ac39fd49ee6d9c6} on Rinkeby test network and it passes compatibility checks by transferring tokens between standard wallets. In terms of gas consumption, its \texttt{transferFrom} function needs at about 37\% more gas than standard \texttt{transferFrom} implementation. We believe this is acceptable for having a secure ERC20 token.


\section{Conclusion}

While this paper is a deep dive into a specific issue with ERC20, it also illustrates a number of higher level lessons for blockchain developers. When ERC20 standard was first implemented, it changed how people used Ethereum, giving rise to an ICO craze with its ease of use~\cite{fenu2018ico}. This led to the deployment of thousands of early implementation of ERC20 tokens which has resulted in numerous attacks on different implementations. Now we see decentralized exchanges relying on existing ERC20 tokens and the \textit{``Multiple Withdrawal Attack''} seems too important to ignore. Fixing existing ERC20 code will help future deployments but cannot fix the already deployed tokens. In addition to deploying secure contracts, we suggest blockchain developers conduct external audits and consider security-by-design practices when dealing with other smart contract implementations. 

\bibliographystyle{unsrt}
\bibliography{references}

\begin{thebibliography}{10}
\footnotesize
\bibitem{Ref00}
Ethereum.
\newblock {Ethereum project repository}.
\newblock \url{https://github.com/ethereum}, May 2014.
\newblock [Online; accessed 10-Nov-2018].

\bibitem{Ref08}
Fabian Vogelsteller and Vitalik Buterin.
\newblock {ERC-20 Token Standard}.
\newblock \url{https://github.com/ethereum/EIPs/blob/master/EIPS/eip-20.md},
  November 2015.
\newblock [Online; accessed 2-Dec-2018].

\bibitem{Ref13}
Mikhail Vladimirov.
\newblock {Attack vector on ERC20 API (approve/transferFrom methods) and
  suggested improvements}.
\newblock
  \url{https://github.com/ethereum/EIPs/issues/20\#issuecomment-263524729},
  November 2016.
\newblock [Online; accessed 18-Dec-2018].

\bibitem{Ref07}
Tom Hale.
\newblock {Resolution on the EIP20 API Approve / TransferFrom multiple
  withdrawal attack \#738}.
\newblock \url{https://github.com/ethereum/EIPs/issues/738}, October 2017.
\newblock [Online; accessed 5-Dec-2018].

\bibitem{eskandari2019sok}
Shayan Eskandari, Seyedehmahsa Moosavi, and Jeremy Clark.
\newblock Sok: Transparent dishonesty: front-running attacks on blockchain.
\newblock {\em International Conference on Financial Cryptography and Data
  Security}, 2019.

\bibitem{Ref10}
OpenZeppelin.
\newblock {openzeppelin-solidity}.
\newblock
  \url{https://github.com/OpenZeppelin/openzeppelin-solidity/blob/master/contracts/token/ERC20/ERC20.sol},
  December 2018.
\newblock [Online; accessed 23-Dec-2018].

\bibitem{Ref11}
ConsenSys.
\newblock {ConsenSys/Tokens}.
\newblock
  \url{https://github.com/ConsenSys/Tokens/blob/fdf687c69d998266a95f15216b1955a4965a0a6d/contracts/eip20/EIP20.sol},
  April 2018.
\newblock [Online; accessed 24-Dec-2018].

\bibitem{Ref06}
Wikipedia.
\newblock {Compare-and-swap}.
\newblock \url{https://en.wikipedia.org/wiki/Compare-and-swap}, July 2018.
\newblock [Online; accessed 10-Dec-2018].

\bibitem{victormeasuring}
Friedhelm Victor and Bianca~Katharina L{\"u}ders.
\newblock Measuring ethereum-based erc20 token networks.
\newblock In {\em International Conference on Financial Cryptography and Data
  Security}, 2019.

\bibitem{Ref23}
Jacques Dafflon, Jordi Baylina, and Thomas Shababi.
\newblock {EIP 777: A New Advanced Token Standard}.
\newblock \url{https://eips.ethereum.org/EIPS/eip-777}, November 2017.
\newblock [Online; accessed 12-Jan-2019].

\bibitem{frowis2018detecting}
Michael Fr{\"o}wis, Andreas Fuchs, and Rainer B{\"o}hme.
\newblock Detecting token systems on ethereum.
\newblock {\em arXiv preprint arXiv:1811.11645}, 2018.

\bibitem{Ref15}
Jordi Baylina, Danil Nemirovsky, and sophiii.
\newblock {minime/contracts/MiniMeToken.sol}.
\newblock
  \url{https://github.com/Giveth/minime/blob/master/contracts/MiniMeToken.sol\#L225},
  December 2017.
\newblock [Online; accessed 23-Dec-2018].

\bibitem{Ref12}
Peter Vessenes.
\newblock {MonolithDAO/token}.
\newblock \url{https://github.com/MonolithDAO/token/blob/master/src/Token.sol},
  April 2017.
\newblock [Online; accessed 23-Dec-2018].

\bibitem{Ref17}
Nate Welch.
\newblock {flygoing/BackwardsCompatibleApprove.sol}.
\newblock
  \url{https://gist.github.com/flygoing/2956f0d3b5e662a44b83b8e4bec6cca6},
  February 2018.
\newblock [Online; accessed 23-Dec-2018].

\bibitem{Ref18}
outofgas.
\newblock {outofgas comment}.
\newblock
  \url{https://github.com/ethereum/EIPs/issues/738\#issuecomment-373935913},
  March 2018.
\newblock [Online; accessed 25-Dec-2018].

\bibitem{Ref03}
M.~Vladimirov and D.~Khovratovich.
\newblock {ERC20 API: An Attack Vector on Approve/TransferFrom Methods}.
\newblock
  \url{https://docs.google.com/document/d/1YLPtQxZu1UAvO9cZ1O2RPXBbT0mooh4DYKjA_jp-RLM/edit#heading=h.m9fhqynw2xvt},
  November 2016.
\newblock [Online; accessed 25-Nov-2018].

\bibitem{Ref16}
Enrique Chavez.
\newblock {StandardToken.sol}.
\newblock
  \url{https://github.com/kindads/erc20-token/blob/40d796627a2edd6387bdeb9df71a8209367a7ee9/contracts/zeppelin-solidity/contracts/token/StandardToken.sol},
  March 2018.
\newblock [Online; accessed 23-Dec-2018].

\bibitem{Ref20}
Dexaran.
\newblock {ERC223 token standard}.
\newblock \url{https://github.com/ethereum/EIPs/issues/223}, March 2017.
\newblock [Online; accessed 12-Jan-2019].

\bibitem{Ref21}
Steve Ellis.
\newblock {transferAndCall Token Standard}.
\newblock \url{https://github.com/ethereum/EIPs/issues/677}, July 2017.
\newblock [Online; accessed 12-Jan-2019].

\bibitem{Ref22}
William Entriken, Dieter Shirley, Jacob Evans, and Nastassia Sachs.
\newblock {ERC-721 Non-Fungible Token Standard}.
\newblock \url{https://github.com/ethereum/EIPs/blob/master/EIPS/eip-721.md},
  January 2018.
\newblock [Online; accessed 12-Jan-2019].

\bibitem{Ref24}
Augusto Lemble.
\newblock {ERC827 Token Standard (ERC20 Extension)}.
\newblock \url{https://github.com/ethereum/eips/issues/827}, January 2018.
\newblock [Online; accessed 12-Jan-2019].

\bibitem{Ref25}
Witek Radomski, Cooke Andrew, Philippe Castonguay, James Therien, and Eric
  Binet.
\newblock {ERC-1155 Multi Token Standard}.
\newblock \url{https://github.com/ethereum/EIPs/blob/master/EIPS/eip-1155.md},
  June 2018.
\newblock [Online; accessed 12-Jan-2019].

\bibitem{Ref26}
Atkins Chang, Noel Bao, Jack Chu, Leo Chou, and Darren Goh.
\newblock {Service-Friendly Token Standard}.
\newblock
  \url{https://github.com/fstnetwork/EIPs/blob/master/EIPS/eip-1376.md},
  September 2018.
\newblock [Online; accessed 12-Jan-2019].

\bibitem{fenu2018ico}
Gianni Fenu, Lodovica Marchesi, Michele Marchesi, and Roberto Tonelli.
\newblock The ico phenomenon and its relationships with ethereum smart contract
  environment.
\newblock In {\em 2018 International Workshop on Blockchain Oriented Software
  Engineering (IWBOSE)}, pages 26--32. IEEE, 2018.

\end{thebibliography}

\end{document}